%% file: main.tex
\documentclass[a4paper, 10 pt, conference]{ieeeconf}  

\IEEEoverridecommandlockouts                              

\usepackage{xcolor}
\usepackage{amsmath} 
\usepackage{amssymb}  
\usepackage{array}
\usepackage{booktabs}
\usepackage{graphicx,nicefrac}
\graphicspath{ {images/} }
\usepackage[normalsize,font=footnotesize]{caption}
\usepackage{xspace}
\usepackage{algorithm}
\usepackage{balance}
\RequirePackage{tikz}

\usetikzlibrary{arrows,calc,positioning,shapes,fit}

\tikzstyle{block} = [rectangle, draw, text centered,
  rounded corners, minimum height=2em]
\tikzstyle{eq} = [circle, draw, minimum height=1.7em, inner sep=1pt] 

\makeatletter
\newcommand{\gettikzxy}[3]{%
  \tikz@scan@one@point\pgfutil@firstofone#1\relax
  \edef#2{\the\pgf@x}%
  \edef#3{\the\pgf@y}%
}
\makeatother

\let\labelindent\relax
\usepackage{enumitem}

\usepackage[style=ieee,backend=bibtex,isbn=false,doi=false,url=false,dashed=false]{biblatex} 
\usepackage{microtype}
\usepackage{geometry}
\usepackage{multirow}
\usepackage{colortbl}

\usepackage{todonotes}

\newcommand{\theProblem}{CF\nobreakdash-EVRP\xspace}
\newcommand{\theAlgorithm}{ComSat\xspace}
\newcommand{\conFreePathSearch}{CFPS\xspace}

\newcommand{\taskSet}{\textrm{$\mathcal{K}$}\xspace}
\newcommand{\nodeSet}{\textrm{$\mathcal{N}$}}
\newcommand{\nodeset}[1]{\ensuremath{\mathcal{N}_#1}\xspace}
\newcommand{\edgeSet}{\textrm{$\mathcal{E}$}}
\newcommand{\Path}{\textrm{$\theta$}}
\newcommand{\EdgeList}{\textrm{$\delta$}}
\newcommand{\Start}{\textrm{$\tau$}\xspace}
\newcommand{\timeHorizon}{\textrm{\emph{T}}}
\newcommand{\serviceTime}{\ensuremath{\mathit{s}}\xspace}

\newcommand{\PairsSet}{\textrm{$\mathcal{P}$}\xspace}
\newcommand{\RoutesSet}{\textrm{$\mathcal{R}$}\xspace}

\newcommand{\PCP}{\emph{Path Changing Problem}\xspace}

\newcommand{\PC}{\emph{Paths\-Changer}\xspace}
\newcommand{\naivePathsChanger}{\ensuremath{\mathit{PC}}\xspace}
\newcommand{\UCGPathsChanger}{\ensuremath{\mathit{GPC}}\xspace}

\newcommand{\CurrentPaths}{\ensuremath{\mathit{CP}}\xspace}
\newcommand{\PreviousPaths}{\ensuremath{\mathit{PP}}\xspace}
\newcommand{\newPaths}{\ensuremath{\mathit{NP}}\xspace}

\newcommand{\useNode}{\textrm{\emph{w}}}
\newcommand{\useEdge}{\textrm{\emph{z}}}

\newcommand{\pathStart}{\textrm{$\xi$}\xspace}
\newcommand{\pathEnd}{\textrm{$\pi$}\xspace}
\newcommand{\incomingEdge}{\textrm{$\mathcal{I}$}}
\newcommand{\outgoingEdge}{\textrm{$\mathcal{O}$}}

\newcommand{\node}{\textrm{\emph{x}}}
\newcommand{\edge}{\textrm{\emph{y}}}
\newcommand{\smallPositive}{\textrm{$\gamma$}\xspace}

\newcommand{\iteration}{\textrm{\emph{h}}\xspace}
\newcommand{\CV}{\emph{CapacityVerifier}\xspace}
\newcommand{\CVP}{\emph{Capacity Verification Problem}\xspace}
\newcommand{\CapacityVerifySolution}{\ensuremath{\mathit{CFS}}\xspace}

\newcommand{\UC}{\emph{Unsat\,Core}\xspace}

\newcommand{\minUC}{\emph{minimal} \UC}
\newcommand{\UnsatCore}{\textrm{$ \mathcal{C}$}\xspace}
\newcommand{\Cbar}[1]{\ensuremath{\,\bar{\!#1}}}
\newcommand{\MUC}{\emph{MUC}\xspace}
\newcommand{\solutionSet}{\textrm{$ \mathcal{S}$}\xspace}
\newcommand{\feasibleSet}{\textrm{$ \mathcal{F}$}\xspace}
\newcommand{\unfeasibleSet}{\textrm{$ \mathcal{U}$}\xspace}

\newtheorem{lemma}{Lemma}
\newtheorem{theorem}{Theorem}
\newtheorem{obs}{Observation}

\title{\LARGE \bf
Leveraging Conflicting Constraints in \\ Solving Vehicle Routing Problems
}

\author{Sabino Francesco Roselli$^{1}$ and Remco Vader$^{2}$ and Martin Fabian$^{1}$ and Knut \AA kesson$^{1}$     
\thanks{We gratefully acknowledge financial support from Chalmers AI Research Centre (CHAIR), ITEA3-projektet AIToC (Artificial Intelligence supported Tool Chain in Manufacturing Engineering), and the Wallenberg AI, Autonomous Systems and Software program (WASP) funded by the Knut and Alice Wallenberg Foundation.
$^{1}$Department of Electrical Engineering, Chalmers University of Technology,
        G\"oteborg, Sweden
        {\tt\small \{rsabino, fabian, knut\}@chalmers.se}.
$^{2}$Department of Mechanical Engineering, Eindhoven University of Technology, Netherlands
        {\tt\small r.m.vader@student.tue.nl}
}}

\begin{document}

\maketitle
\thispagestyle{empty}
\pagestyle{empty} 

\begin{abstract}

The Conflict-Free Electric Vehicle Routing Problem (\theProblem) is  a combinatorial optimization problem of designing routes for vehicles to visit customers such that a cost function, typically the number of vehicles or the total travelled distance, is minimized. The \theProblem involves constraints such as time windows on the delivery to the customers, limited operating range of the vehicles, and limited capacity on the number of vehicles that a road segment can simultaneously accommodate.
In previous work, the compositional algorithm \emph{\theAlgorithm} was introduced and that solves the \theProblem by breaking it down into sub-problems and iteratively solve them to build an overall solution.
Though \theAlgorithm showed good performance in general, some problems took significant time to solve due to the high number of iterations required to find solutions that satisfy the road segments' capacity constraints. 
The bottleneck is the \PCP, i.e., the sub-problem of finding a new set of shortest paths to connect a subset of the customers, disregarding previously found shortest paths. This paper presents an improved version of the \PC function to solve the \PCP that exploits the \emph{unsatisfiable core}, i.e., information on which constraints conflict, to guide the search for feasible solutions. Experiments show faster convergence to feasible solutions compared to the previous version of \PC.

\end{abstract}

\section{Introduction} \label{sec:intro}

We consider scheduling a fleet of mobile robots, in the sequel referred to as Automated Guided Vehicles (AGVs), that pick-up and deliver components to workstations within specified time-windows. The AGVs move on a predefined road network, where each road segment has a maximum number of AGVs it can accommodate at a specific time. The problem is motivated by an industrial need to develop more flexible logistic systems to deliver components just-in-time to an assembly line.
 
In this scenario, in addition to time-windows in which the components should be delivered, a scheduler needs to consider additional constraints. First, AGVs have a limited operating range and need to recharge their battery when the state-of-charge becomes low. Second, jobs have specific requirements on the AGV eligible to execute them. 
Finally, the number of AGVs on road segments and workstations are limited to allow low-level trajectory planning problems to be feasible.
Thus, we define the \emph{capacity} of the road segments, intersections, and workstations and include \emph{capacity constraints}.
A schedule is said to be  \emph{conflict-free} if it fulfills the capacity constraints at all times.

The problem of computing conflict-free routes was first introduced in \cite{krishnamurthy1993developing} and tackled by means of column generation. In \cite{correa2007scheduling}, conflict-free routing in combination with scheduling of jobs for flexible manufacturing systems is discussed.
An ant colony algorithm is applied to the problem of job shop scheduling and conflict free routing of AGVs by~\cite{saidi2015ant}.
In~\cite{yuan2016research}, a collision-free path planning for multi AGV systems based on the $A^*$ algorithm is presented. 
Another heuristic approach to solve the conflict-free routing problem with storage allocation is presented by~\cite{thanos2019dispatch}. 
In~\cite{murakami2020time}, a MILP formulation to design conflict-free routes for capacitated vehicles is presented.
In~\cite{zhong2020multi} is presented a hybrid evolutionary algorithm to deal with conflict-free AGV scheduling in automated container terminals, 
and~\cite{chen2021integrated} handles the problem of conflict-free routing of AGVs by a meta-heuristic improvement strategy based on large neighbourhood search. 
Hence, conflict-free routing and scheduling has been addressed previously, but to the best of our knowledge, there is no work in the literature that tackles all above mentioned constraints at once.
Therefore,~\cite{roselli2021solving} introduced the \emph{Conflict-Free Electric Vehicle Routing Problem} (\theProblem). The \theProblem  is an extension of the vehicle routing problem (VRP) \cite{dantzig1959truck}, involving the additional constraints.
In ~\cite{roselli2022compo_algo} a compositional algorithm, \theAlgorithm, for solving the \theProblem is proposed. \theAlgorithm breaks down \theProblem into sub-problems and iteratively solves these to find a feasible solution to the overall problem. Experimental and analytical evaluation shows that \theAlgorithm generates high-quality but not necessarily optimal solutions. 
Briefly, \theAlgorithm computes routes to serve the customers, and assigns vehicles to the routes attempting to make the execution of the system conflict-free.
In a plant there can be several ways to travel from one customer's location to another.
Initially, \theAlgorithm uses the shortest paths among the customers' locations when designing the routes.
However, if a feasible schedule cannot be achieved using the shortest paths, alternative paths have to be found, which is handled by the \emph{Conflict-free Paths Search} (\conFreePathSearch). \conFreePathSearch  is composed of two main functions;
the \PC function, that finds alternative sets of paths if the current schedule violates the capacity constraints, and the \CV function, that checks whether the schedule is conflict-free or not.

Experiments show that when a solution computed using the shortest paths violates the capacity constraints, finding alternative paths using the \PC function may require multiple iterations. This does not come unexpected, since the number of possible paths in a graph can be high, and minimizing the cumulative length while looking for alternative paths does not guarantee that the schedule will be conflict-free. In this paper we focus on the \conFreePathSearch and present improved versions of the \PC and \CV that, in many cases, find feasible solutions faster.

The sub-problems in \theAlgorithm are modelled as Satisfiability Modulo Theory (SMT) problems~\cite{barret_2009,DeMoura_2011}, as SMT solvers have shown to be efficient in solving combinatorial problems~\cite{weber2019smt}.

Moreover, some SMT solvers come with algorithms that allow them to deal with optimization problems~\cite{sebastiani2020optimathsat}. Two sub-problems in \theAlgorithm, marked by the round boxes in Fig.~\ref{fig:Comsat_flowchart} (see below) are optimization problems. 

For the \conFreePathSearch polynomial time algorithms exist to find paths in graphs,~\cite{gross2003handbook}. However, modelling the \PCP as an SMT problem is beneficial as it allows to define problem-specific requirements, such as not returning solutions that are already proven infeasible because they violate the capacity constraints. Moreover, when a problem is infeasible, SMT solvers have the ability to return a \emph{Minimal Unsatisfiable Core} (\MUC)~\cite{cimatti2011computing}, i.e., one of the (possibly many) smallest subsets of constraints that make the problem infeasible. The \MUC can provide useful information about why a problem is infeasible and can therefore be used to guide the search towards a feasible solution~\cite{selsam2019guiding}. 

When dealing with the \theProblem, the \MUC can be extracted when the \CVP is infeasible and used to define additional constraints for the \PCP, to increase the chances of finding a feasible schedule.

The contributions in this paper are: (i) exploitation of SMT solvers' \MUC to extract information about the infeasibility of an SMT formula representing a conflicting schedule for a VRP; (ii) use of such information to find conflict-free schedules; (iii) performance comparison between the unguided and \MUC guided paths search over a set of \theProblem problem instances.

The remainder of the paper is organized as follows. Preliminaries are presented in Section~\ref{sec:problem_formulation}. Section~\ref{sec:math_model} presents the mathematical models of the sub-problems that form the \conFreePathSearch and how it is improved using the \MUC from the \CVP.
Proof of soundness and completeness of the procedure is provided in Section~\ref{sec:proofs}. In Section~\ref{sec:experiments}, the results of the analysis over a set of problem instances are presented. Finally, conclusions are drawn in Section~\ref{sec:conclusions}.

\section{Preliminaries} \label{sec:problem_formulation}

In the \theProblem the plant layout is represented by a finite, strongly connected, weighted, directed graph, where edges represent road segments and nodes represent either intersections between road segments or customers' locations.
A customer is defined by a unique (numerical) identifier, a location, and a time window, i.e., a lower and upper bound that represent the earliest and latest arrival time allowed to serve the customer.
Edges have two attributes, the first representing the road segment's length, and the second its capacity. The capacity is $2$ if two vehicles can simultaneously travel in opposite directions, $1$ otherwise.

The following definitions are provided:
\begin{itemize}

    \item \emph{Node}: a location in the plant. 
    A node can only accommodate one vehicle at a time unless it is a \emph{hub} node that can accommodate an arbitrary number of vehicles.
    \begin{itemize}[label={}]
        \item \nodeSet: a finite set of nodes.
        \item $\nodeset{H} \subseteq \nodeSet $: the set of hub nodes.
    \end{itemize}
    
    \item \emph{Edge}: a road segment that connects two nodes.
    \begin{itemize}[label={}]
        \item $\edgeSet \subseteq \nodeSet \times \nodeSet$: the finite set of direct edges.
        \item $\bar{e}$: the reverse edge of edge $e \in \edgeSet$.
        \item $ d_e \in \mathbb{R}_+$: the length of edge $e \in \edgeSet$.
        \item $ g_e \in \{1,2\}$: the capacity of edge $e \in \edgeSet$.
    \end{itemize}
    \item \emph{Time horizon}: a fixed, continuous point of time when all jobs have ended, assuming they start at time $0$.
    \begin{itemize}[label={}]
        \item $ \timeHorizon $: the time horizon. 
    \end{itemize}
    
    \item \emph{Customer}: Entity representing a task to be executed by a vehicle, e.g., a pickup or delivery of material, that needs to be visited exactly once by the vehicle. A customer is always associated with a node where the pickup/delivery operation is executed, and has a time window indicating the earliest and latest time at which it can be visited. Unless explicitly given, the time window is the entire time span $[0,T]$. 
 
    \begin{itemize}[label={},leftmargin=*]
    \item Let $\taskSet$ be the finite set of all customers, and let 
    \item $ l_{k}, u_k \in \mathbb{R}_+, \ k \in \taskSet$ be the time window's lower ($l_k$) and upper ($u_k$) bound for customer $k$ such that $u_{k} > l_{k}$.
    \item  Also let $ \serviceTime_{k} \in \mathbb{R}_+$ and $L_{k} \in \nodeSet$, for $k \in \taskSet$, be the service time and location of customer $k$, respectively. 
    \end{itemize}
    \item \emph{Route}: an ordered set of unique customers.
    \begin{itemize}[label={}]
        \item $r_j = \langle k_{j1},\ldots,k_{jm} \rangle,\ m
        \leq |\taskSet|,\, k_{ji}\in\taskSet,$
        \item $\qquad i=1,\ldots,m,\, k_{jl}\neq k_{ji} \textrm{ for } i\neq l$.        
    \end{itemize}
    A route can at most include all customers, therefore $m \leq |\taskSet|$.
    \item \emph{Route set}: a set of routes such that each customer belongs to exactly one route, thus guaranteeing that all customers are served.
    \begin{itemize}[label={}]
        \item $ \RoutesSet = \{ r_1,\ldots,r_m \},\  m \leq |\taskSet|  $
    \end{itemize}
    A route contains at least one customer, hence $m \leq |\taskSet|$.
    \item \emph{Route start}: the starting time $\Start_r$ of route $r$, computed by the function \emph{Assign}. $\Gamma$ is the set that contains the \emph{route start} of each route. 
    \begin{itemize}[label={}]
        \item $\Gamma = \{\Start_r \in \mathbb{R} \, |\, r \in \RoutesSet \} $
    \end{itemize}

    \item \emph{Pair Set of route $r$}: set containing the sequence of customers of a route $r=\langle k_{1},\ldots,k_{m} \rangle$, grouped as pairs in sequence.
    \begin{itemize}[label={}]
        \item $\PairsSet_r = \{ \langle k_1,k_2\rangle,\langle k_2,k_3 \rangle,\ldots,\langle k_{m-1},k_m \rangle \} $
        
    \end{itemize}
    \item \emph{Path}: ordered set of unique nodes. It is used to keep track of how vehicles are travelling among customers of routes, since  each pair of customers in a route is connected by a path.
    \begin{itemize}[label={}]
        \item $\Path_p = \langle n_1,\ldots,n_m \rangle, \ p \in \PairsSet_r, \ m \leq |\nodeSet |,$
        \item $\qquad n_i\in\nodeSet,\, i = 1,\ldots,m $
    \end{itemize}
    \item \emph{Edge sequence}: ordered set of unique edges for a given path $\Path_p$. 
    \begin{itemize}[label={}]
        \item $\EdgeList_p = \langle e_1,\ldots,e_m \rangle, \ p \in \PairsSet_r, \ m = |\Path_p | - 1,$
        \item $ \qquad e_i\in\edgeSet,\, i = 1,\ldots,m $
    \end{itemize}
\end{itemize}

In order to clarify which part of \theAlgorithm is analyzed and improved in this work, let us recap briefly how the algorithm works. Fig.~\ref{fig:Comsat_flowchart} shows a simplified flowchart of \theAlgorithm that illustrate the concepts of this paper. The first step of \theAlgorithm is to design a set of routes $\RoutesSet$ to serve all the customers; at this point, the shortest path between any two customers is computed using Dijkstra's algorithm \cite{dijkstra1959note}.
\newpage
\begin{figure}[h]
    \centering
    \scalebox{0.87}{\input{ComSat_Flowchart}}
    \caption{Flowchart of \theAlgorithm.}
    \label{fig:Comsat_flowchart}
\end{figure}

This optimization problem is handled by the function \emph{Router} and must guarantee that the routes meet specific requirements such as maximum length, specific ordering among the customers and time windows. If this step is infeasible the \theProblem instance has no solution and the algorithm terminates. If this step is feasible, the function \emph{Assign} will try to allocate available vehicles to the routes and compute a start time $\tau_r, \ \forall r \in \RoutesSet$, to the routes. If this step is infeasible then \emph{Router} will try to find different routes, but if it is feasible, the \emph{CapacityVerifier} checks if the current set of routes is conflict-free. More details on the functions \emph{Router} and \emph{Assign} can be found in~\cite{roselli2022compo_algo}.

\subsection{\emph{The} \minUC}

For infeasible problems, there can be identified a subset of the constraints that conflict, meaning they cannot all simultaneously be satisified. Such a subset is called an \UC. An \UC with the property that removing any one of the constraints makes the \UC feasible, is said to be \emph{minimal}.

Formally, given an SMT formula $\varphi$ and set of conflicting constraints $\UnsatCore\subseteq\varphi$, $\UnsatCore$ is a \MUC of $\varphi$ if removing any constraint  $\UnsatCore_i\in\UnsatCore$ makes $\UnsatCore\setminus\UnsatCore_i$ no longer infeasible; removing $\UnsatCore$ removes the particular conflict represented by the \MUC. 
Consequently, for an infeasible problem with a \MUC $\UnsatCore$, adding to the problem a constraint that prevents all the constraints in $\UnsatCore$ to be simultaneously active will resolve this particular conflict.

The na\"ive approach to \MUC extraction, \cite{dershowitz2006scalable}, successively removes constraints and solves the problem again; if the problem is still infeasible after a constraint has been removed that constraint does not belong to a \MUC.
There exist more efficient approaches though; the \MUC~\cite{huang2005mup} algorithm based on efficient manipulation of \emph{Binary Decision Trees} guarantees the extraction of a \minUC. \cite{nadel2010boosting} presents an algorithm based on the resolution graph \cite{kroening2016decision} for \MUC extraction. \cite{nadel2013efficient} improves the resolution based algorithm using \emph{model rotation} and \emph{path strengthening}.

\section{The Conflict-free Paths Search} \label{sec:math_model}

In this section the two sub-problems that form the \conFreePathSearch are presented. The \CVP is modelled as a job shop problem (JSP), in order to exploit the good performance of the SMT solver Z3~\cite{bjorner2015nuz} in dealing with JSPs, as demonstrated in~\cite{roselli2018smt}. The model formulation for the \PCP is inspired by~\cite{aloul2006identifying}.

The following logical operators are used as a shorthand to express cardinality constraints \cite{sinz2005towards} in the sub-problems:
\begin{itemize}
    \item[] $\textrm{EN}(A,n):$ exactly $n$ variables in the set $A$ are true;
    \item[] $\textrm{If}(c,o_1,o_2):$ if $c$ is \emph{true} returns $o_1$, else returns $o_2$.
\end{itemize}
We will write $\textrm{EN}_{m\in M}(m,n)$  to denote $ \textrm{EN}({\bigcup\limits_{m\in M}\{m\},n)}$ in order to shorten the notation.

\subsection{\emph{The} \CVP}

The \CVP aims to find a feasible schedule for the vehicles, where the routes that the vehicles are assigned to satisfy the capacity constraints of the edges. 

In this work the \CVP, as defined in \cite{roselli2022compo_algo}, has been extended to account for pairs as well, since the information about conflicts must be related to a specific pair to define additional constraints in the \PC. 

Let $n_\textrm{rpe}$ be the node visited before edge $e$ of pair $p$ of route $r$, and let $e_\textrm{rpn}$ be the node visited before node $n$ on pair $p$ of route $r$.
Similarly, let $n^\textrm{rpe}$ be the node visited after edge $e$ of pair $p$ of route $r$, and let $e^\textrm{rpn}$ be the edge visited after node $n$ on pair $p$ route $r$. Let $p_r^0$ be the first pair of route $r$ and $n^*_r$ be its starting node.

\subsubsection*{Example of Routes, Pairs, Nodes, and Edges} \hfill\\
Let $\taskSet= \{k_1,\ldots,k_7\}$ and $\nodeSet=\{n_1,\ldots,n_{20}\}$. Let $L_{k_1} = n_1 $ and $ L_{k_2} = n_7 $, and assume two routes  designed to serve all customers:
$r_1 = \langle k_1, k_2, k_5, k_7 \rangle, \ r_2 = \langle k_3, k_4, k_6 \rangle $.

In order to clarify the notation introduced above, let us analyze $r_1$. First, the set of pairs for $r_1$ is defined as \\
\indent $\PairsSet_{r_1} = \{ \langle k_1,k_2 \rangle,\langle k_2,k_5 \rangle,\langle k_5,k_7 \rangle \} $. \\
Then, let us assume that the \emph{path} and \emph{edge sequence} for pair $\langle k_1,k_2 \rangle$ are the following: \\
\indent $\Path_{\langle k_1,k_2 \rangle} = \langle n_1,n_2,n_4,n_5,n_7 \rangle $, \\
\indent $\EdgeList_{\langle k_1,k_2 \rangle} = \langle \langle n_1,n_2 \rangle,\langle n_2,n_4 \rangle,\langle n_4,n_5 \rangle,\langle n_5,n_7 \rangle \rangle $.\\
Then $p_{r_1}^0 = \langle k_1,k_2 \rangle$ and $n^*_{r_1} = n_1$. 
Also, let $p = \langle k_1,k_2 \rangle$; then for $e = \langle n_1,n_2 \rangle$, $n_{r_1pe} = n_1$, and $n^{r_1pe} = n_2$; 
for $n = n_1$, $e^{r_1pn} = \langle n_1,n_2 \rangle$, and
for $n = n_2$, $e_{r_1pn} = \langle n_1,n_2 \rangle$.

For each node it must also be specified whether there exists a time window, since some of the nodes are only intersections of road segments in the real plant, while others are actual customers.
Let $l_{rpn}$ and $u_{rpn}$ be the earliest and latest arrival time, respectively, at node $n$ of pair $p$ of route $r$;
let $\serviceTime_{rpn}$ be the service time at node $n$ of pair $p$ of route $r$. Finally, let \smallPositive $> 0$ be a small real constant used to prevent \emph{swapping} of vehicles' positions between a node and the previous or following edge.

The \emph{Capacity Verification Problem} decision variables are:
\begin{itemize}[label={}]
    \item $\node_{rpn}$: non-negative real variable that models when a vehicle executing route $r$ starts using node $n$ in pair $p$;
    \item $\edge_{rpe}$: non-negative real variable that models when a vehicle executing route $r$ starts using edge $e$ in pair $p$;
\end{itemize}
The model for the \emph{Capacity Verification Problem} is:
\begin{flalign}
    &\node_{rp_r^0n^*_r} \geq \Start_r,  \ \forall r \in \RoutesSet \label{eq:route_start} \\
    &\edge_{rpe} \geq \node_{rpn_{rpe}} + \serviceTime_{rpn_{rpe}}, \ \forall r \in \RoutesSet, \ p \in \PairsSet_r, \ e \in \EdgeList_p \label{eq:visit_precedence_1} \\
    &\node_{rpn} = \edge_{rpe_{rpn}} + d_{e_{rpn}}, \ \forall r \in  \RoutesSet, \ p \in \PairsSet_r, \ n \in \Path_p \label{eq:visit_precedence_2} \\
    &\node_{rpn} \geq l_{rpn} \wedge \node_{rpn} \leq  u_{rpn}, \nonumber \\
    &\quad\forall r \in  \RoutesSet, \ p \in \PairsSet_r, \ n \in \Path_p \label{eq:visit_tw} \\
    &\node_{r_1p_1n} \geq \edge_{r_2p_2e^{r_1p_1n}} + \smallPositive \ \vee \ \node_{r_2p_2n} \geq \edge_{r_1p_1e^{r_2p_2n}} + \smallPositive,  \nonumber \\
    & \quad\forall r_1,r_2 \in \RoutesSet, \  r_1  \neq r_2, \ p_1 \in \PairsSet_{r_1}, \ p_2 \in \PairsSet_{r_2} \nonumber \\
    & \quad n \in \Path_{p1} \cap \Path_{p2}, \ n \notin \nodeset{H} \label{eq:nodes_no_swap}\\
    &\edge_{r_1p_1e} \geq \edge_{r_2p_2e} + \smallPositive  \vee   \edge_{r_2p_2e} \geq \edge_{r_1p_1e} + \smallPositive, \nonumber \\
    &\quad \forall r_1,r_2 \in \RoutesSet, \, r_1  \neq r_2, \ p_1 \in \PairsSet_{r_1}, \ p_2 \in \PairsSet_{r_2}, \nonumber \\
    & \quad e \in \EdgeList_{p_1} \cap \EdgeList_{p _2} \label{eq:edges_direct}\\ 
    &\edge_{r_1p_1e_1} \geq \edge_{r_2p_2e_2} +  d_{e_2} \ \vee \edge_{r_2p_2e_2} \geq  \edge_{r_1p_2e_1} + d_{e_1}, \nonumber \\
    & \quad  \forall r_1,r_2 \in \RoutesSet, \ r_1 \neq r_2, \ p_1 \in \PairsSet_{r_1}, \ p_2 \in \PairsSet_{r_2},\nonumber \\
    & \quad e_1 \in \EdgeList_{p_1}, \ e_2 \in \EdgeList_{p_2}, \ e_1 = \bar{e}_2, \ g_{e_1} = g_{e_2} = 1  \label{eq:edges_inverse}
\end{flalign}

\noindent
\eqref{eq:route_start} constrains the start time of a route; \eqref{eq:visit_precedence_1} and \eqref{eq:visit_precedence_2} define the precedence among nodes and edges to visit in a route; \eqref{eq:visit_tw} enforces time windows on the nodes that correspond to the customers; \eqref{eq:nodes_no_swap} prevents vehicles from using the same node at the same time; \eqref{eq:edges_direct} and \eqref{eq:edges_inverse} constrain the transit of vehicles over the same edge. If two vehicles are using the same edge from the same node, one has to start at least $\gamma$ after the other and if two vehicles are using the same edge from opposite nodes, one has to fully transit before the other one can start. 

Based on the model described above, the algorithm \CV (\emph{CV}) is defined, that takes a set of routes $\RoutesSet$, the start times in $\Gamma$, and the current set of paths \CurrentPaths as input and returns:
\begin{itemize}
    \item \CapacityVerifySolution, a list that expresses where each vehicle is at each time; this is empty if the problem is infeasible.
    \item \Cbar{\UnsatCore}, the \UC relative to constraints \eqref{eq:nodes_no_swap}-\eqref{eq:edges_inverse} (see  Section~\ref{sec:improved_paths_search}); this is empty if the problem is feasible. 
\end{itemize}

\subsection{Paths Changing Problem}

In the \emph{Paths Changing Problem}, alternative paths are computed to connect the consecutive customers of each route. Finding alternative paths may be necessary when, for a given set of routes \RoutesSet and starting times $\Gamma$, no feasible schedule exists. The \emph{Capacity Verification Problem} may be infeasible due to the current set of paths that connect the customers' locations, therefore a different set may lead to a feasible solution.
A route is defined as a sequence of customers, and for any two consecutive customers there is a path (a sequence of edges) connecting them. Therefore, for a route containing $i + 1$ customers we will have $i$ paths and for each path we can define a start and an end node, $\pathStart_i$ and $\pathEnd_i$, respectively.
The sets of outgoing and incoming edges for a certain node $n$ are denoted $\outgoingEdge_n$ and $\incomingEdge_n$, respectively.

Decision variables used to build the model are:
\begin{itemize}[label={}]
    \item $\useNode_{rpn}$: Boolean variable that represents whether the pair $p$ of route $r$ is using node $n$;
    \item $\useEdge_{rpe}$: Boolean variable that represents whether the pair $p$ of route $r$ is using edge $e$;
\end{itemize}

This problem can be split into $ r \cdot i$ sub-problems (assuming all routes have $i + 1$ customers) that find paths for each route separately; simpler and smaller models are faster to solve. Unfortunately it may be necessary to explore different combinations of paths, so to retain the information we have only one model. Therefore, let the optimal solution to the \emph{Path Changing Problem} found at iteration \iteration be 
$$ \CurrentPaths = \bigcup_{ \substack{ r \in \RoutesSet \\ p \in \PairsSet_r \\ e \in \edgeSet } } { \{ \useEdge^*_{rpe} \}, } $$
where $z^*_{rpe}$ is the value of $z_{rpe}$ in the current solution; also, let \PreviousPaths be the set containing the optimal solutions found until the $(\iteration-1)$-th iteration.
The model is then:
\begin{flalign}
    & \min_{ r \in \RoutesSet, \ p \in \PairsSet_r, \ n \in \edgeSet} \sum{\textrm{If}(\useEdge_{rpe},d_e,0)} \label{eq:path_changing_cost_function} \\
    & \useNode_{rp\pathStart_p} \wedge \useNode_{ri\pathEnd_p}, \qquad \qquad \quad \ \ \forall p \in \PairsSet_r, \ r \in \RoutesSet \label{eq:start_and_end_are_true} \\
    & \textrm{EN}_{e \in \outgoingEdge_{\pathStart_p} }{(\useEdge_{rpe},1)}, \qquad \qquad \ \forall p \in \PairsSet_r, \ r \in \RoutesSet \label{eq:only_one_edge_for_start}\\
    & \textrm{EN}_{e \in \incomingEdge_{\pathStart_p} }{(\useEdge_{rpe},1)}, \qquad \qquad \ \forall p \in \PairsSet_r, \ r \in \RoutesSet \label{eq:only_one_edge_for_end}\\
    & \useEdge_{rpe} \implies \neg{\useEdge_{rp\bar{e}}}, \quad \ \ \forall p \in \PairsSet_r, \ r \in \RoutesSet, \ e \in \edgeSet \label{eq:not_both_directions}\\
    & \bigwedge_{n \in \nodeSet, n \neq \pathStart_p, n \neq \pathEnd_p}\textrm{If}(\useNode_{rpn}, \nonumber \\
    & \qquad \qquad \textrm{EN}_{e \in \outgoingEdge_n}{(\useEdge_{rpe}},1) \wedge \textrm{EN}_{e \in \incomingEdge_n}{(\useEdge_{rpe},1)}, \nonumber \\
    & \qquad \qquad \qquad   \textrm{EN}_{e \in \outgoingEdge_n}{(\useEdge_{rpe},0)} \wedge \textrm{EN}_{e \in             \incomingEdge_n}{(\useEdge_{rpe},0)}), \qquad \nonumber \\
    & \qquad \qquad \qquad \qquad \qquad \quad \ \ \forall p \in \PairsSet_r, \ r \in \RoutesSet \label{eq:exactly_two_edges}\\
    &\bigvee_{\useEdge_{rpe} \in \CurrentPaths}{\neg{\useEdge_{rpe}}}, \qquad \qquad \qquad \qquad   \forall \CurrentPaths \in \PreviousPaths  \label{eq:previous_paths}
\end{flalign}

The cost function \eqref{eq:path_changing_cost_function} to minimize is the cumulative length of the used edges; \eqref{eq:start_and_end_are_true} guarantees that, for each path of each route, the start and end nodes are used; \eqref{eq:only_one_edge_for_start} and \eqref{eq:only_one_edge_for_end} make sure that exactly one outgoing (incoming) edge is incident with the start (end) node of a route; \eqref{eq:not_both_directions} makes sure that a path is not allowed to use both an edge and its reverse; \eqref{eq:exactly_two_edges} guarantees that if a node (different from the start or end) is selected, exactly one of its outgoing and one of its incoming edges will be used. On the other hand, if a  node is not used, none of its incident edges will be used; finally, \eqref{eq:previous_paths} rules out all the previously found solutions.

Based on the model described above the function \PC (\naivePathsChanger) is defined, that takes the previous paths \PreviousPaths as input and returns
a new set of paths \newPaths. If the \emph{Paths Changing Problem} is infeasible then $\newPaths=\emptyset$.

Up to this point, unless specified otherwise, the models presented are taken from \cite{roselli2022compo_algo}.

\subsection{Exploiting the \MUC} \label{sec:improved_paths_search}

Experiments reported in~\cite{roselli2022compo_algo}, show that \theAlgorithm performs well for many problem instances, however, for some specific instances \theAlgorithm failed to find feasible solutions in reasonable time. Investigations revealed the \naivePathsChanger to be the culprit. The reason is that it searches blindly through the possible paths that connect any two customers, while minimizing the paths' cumulative length.
A \emph{conflict-free} solution may involve paths that are quite longer than the current ones though, and the \naivePathsChanger will have to explore many \emph{shorter} solutions before finding the right one. 
Improving the performance of the \PC would be beneficial for the overall performance of \theAlgorithm, and letting the \MUC guide the paths changing is such an improvement.

When extracting the \MUC, it is possible to only track specific constraints. This feature can be exploited to focus only on the capacity constraints violations. In fact, since time windows and service time are not flexible, it is of little to no use to track constraints represented by \eqref{eq:route_start}-\eqref{eq:visit_tw}.
Also, an infeasible formula $\varphi$ may have multiple \emph{MUCs}; in the \theProblem this means that conflicts may arise at different locations in the plant. In order to catch all of them, it is possible to iteratively relax the conflicting constraints from the initial formula and solve it again, until it becomes feasible. The formula will indeed become feasible eventually, since it is based on a feasible solution \RoutesSet and only the capacity constraints can make it infeasible; in the worst case all such constraints will be removed during the iterations.
Note that, since not all constraints are tracked, the set of constraints $\Cbar{\UnsatCore}$ returned is not an actual \UC, since $\Cbar{\UnsatCore}$ would only make the problem infeasible in conjunction with the untracked constraints. Nonetheless, it provides the information about the conflicts needed to guide the search of paths. 

Let $\varphi_0$ be the conjunction of constraints \eqref{eq:route_start}-\eqref{eq:edges_inverse}. Assume that $\varphi_0$ is infeasible, and let $\Cbar{\UnsatCore}_0$ be the subset of a \MUC retrieved by tracking constraints \eqref{eq:nodes_no_swap}-\eqref{eq:edges_inverse}. Then let $\varphi_1 = \varphi_0 \setminus \Cbar{\UnsatCore}_0$, also infeasible, and let $\Cbar{\UnsatCore}_1$ be the subset of a \MUC retrieved by tracking constraints defined by \eqref{eq:nodes_no_swap}-\eqref{eq:edges_inverse}, not including the ones in $\Cbar{\UnsatCore}_0$. 
In general, the constraints in $\Cbar{\UnsatCore}_{i-1}$ can be iteratively relaxed to obtain a new formula $\varphi_i$, until a feasible $\varphi_n = \varphi_0 \setminus ( \Cbar{\UnsatCore}_0 \cup \ldots \cup \Cbar{\UnsatCore}_{n-1} )$ is found.
Then $\Cbar{\UnsatCore} =   \Cbar{\UnsatCore}_0 \cup \ldots \cup \Cbar{\UnsatCore}_{n-1} $ contains all the conflicts due to the capacity constraints.

Each constraint represented by \eqref{eq:nodes_no_swap}-\eqref{eq:edges_inverse} is defined over two routes $r_1$ and $r_2$ and their pairs $p_1$ and $p_2$ for a specific node $n$ or edge $e$; therefore, if the constraint is part of $\Cbar{\UnsatCore}$, the routes and pairs that caused the conflict over $n$ or $e$ can be identified. 
If the conflict was generated by a constraint from \eqref{eq:nodes_no_swap}, then the following constraint is added to \eqref{eq:path_changing_cost_function}-\eqref{eq:previous_paths}:
\begin{flalign}
    & \neg( \useNode_{r_1p_1n}) \vee \neg( \useNode_{r_2p_2n}). \label{eq:avoid_node}
\end{flalign}
On the other hand, if the conflict was caused by  constraint from \eqref{eq:edges_direct} or \eqref{eq:edges_inverse}, the following constraint is added to  \eqref{eq:path_changing_cost_function}-\eqref{eq:previous_paths}:
\begin{flalign}
    & \neg( \useEdge_{r_1p_1e}) \vee \neg( \useEdge_{r_2p_2e}).    
    \label{eq:avoid_edge} 
\end{flalign}

Constraints \eqref{eq:avoid_node} and \eqref{eq:avoid_edge} force at least one of the routes involved in the conflict to avoid the specific node (edge, respectively) when computing a path for the pairs involved in the conflict. The constraint is formulated so that the choice of the route to change is left to the solver, including the possibility of changing both routes; since the problem is an optimization, the solver will choose the change that leads to the shortest cumulative paths length.

Based on the model described by~\eqref{eq:path_changing_cost_function}-\eqref{eq:avoid_edge},  the function \emph{MUC-Guided-Paths-Changer} (\UCGPathsChanger) is defined, that takes the previous paths \PreviousPaths and $\Cbar{\UnsatCore}$ as input and returns
a new set of paths \newPaths. If the \PCP is infeasible $\newPaths=\emptyset$.

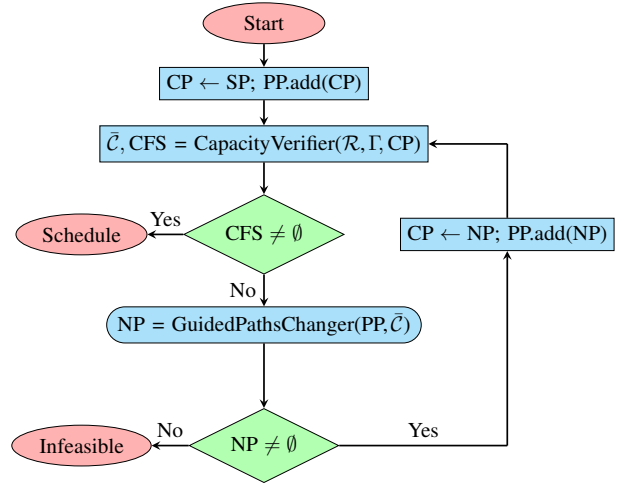
\begin{figure}[h]
    \centering
    \scalebox{0.80}{\input{CFPS_Flowchart}}
    \caption{Flowchart of the \MUC-Guided-\conFreePathSearch.}
    \label{fig:flowchart}
\end{figure}

Since for each constraint in $\Cbar{\UnsatCore}$ a new constraint is added to the \UCGPathsChanger, it is imperative that the \UC returned when the \emph{CV} is infeasible is \emph{minimal}. This is so because if the \UC is not minimal, it could contain constraints that are not actually causing \emph{capacity conflicts}. These constraints would in turn lead to defining constraints \eqref{eq:avoid_node} and \eqref{eq:avoid_edge} in the \UCGPathsChanger that may remove feasible solutions.

Fig.~\ref{fig:flowchart} summarizes the steps required to find a conflict-free schedule \CapacityVerifySolution, if such exists, using the improved paths searching algorithm \UCGPathsChanger. As mentioned, it is assumed that routes \RoutesSet and their start times $\Gamma$ have already been computed. The shortest paths between any two customers are computed using Dijkstra's algorithm and then set as the current paths \CurrentPaths to travel among customers. Also, \CurrentPaths are added to the list of previous paths \PreviousPaths.

Then the \emph{CV} will check such routes against the capacity constraints; if this sub-problem has a feasible solution the algorithm terminates and a conflict-free schedule is returned. Otherwise $\Cbar{\UnsatCore}$ is extracted as described in the previous paragraph and the the \UCGPathsChanger algorithm is invoked. \UCGPathsChanger will use the information about previously computed paths \PreviousPaths and the information about conflicts from $\Cbar{\UnsatCore}$ to compute new paths \newPaths, which will be set as the current paths and stored in \PreviousPaths. At this point the \emph{CV} is run again using the new paths. The iterations between the two algorithms continue until either the \emph{CV} is feasible, or the \UCGPathsChanger is infeasible, i.e., there are no feasible, conflict-free paths to execute the routes \RoutesSet with the start times $\Gamma$.

\section{Proof of Soundness and Completeness} \label{sec:proofs}

In this section, proof of soundness and completeness of the \UC Guided \conFreePathSearch is provided. The underlying idea for the proof is the following. There exists a finite number of solutions to the \PCP; the $\UCGPathsChanger$ can enumerate at least all feasible solutions to the \PCP; if a solution that satisfies the \emph{Capacity Constraints} does exists, the $\UCGPathsChanger$ will eventually find it, otherwise it will declare the problem infeasible. 

Let \solutionSet be the set of possible solutions to a \PCP; let us divide \solutionSet into the set of conflict-free solutions \feasibleSet and the set of conflicting solutions \unfeasibleSet. In other words a solution to the \PCP from \feasibleSet will make the \CVP feasible, while a solution from \unfeasibleSet will not. If the \conFreePathSearch is infeasible, then $\solutionSet = \unfeasibleSet$ and $\feasibleSet = \emptyset$. In this case, even if the $\UCGPathsChanger$ is not able to find all feasible solutions \feasibleSet, there is none to find. 

In case the \conFreePathSearch is feasible though, in order to prove completeness it is necessary to guarantee that at least all feasible solutions \feasibleSet can be found by \UCGPathsChanger. This is proven for the \naivePathsChanger, since each call of the \naivePathsChanger function will find the next optimal solution to the \PCP, whether it belongs to \feasibleSet or not, until all solutions are enumerated. However in the $\UCGPathsChanger$ there are additional constraints that may remove feasible solutions. In the proof it is shown that such additional constraints only remove infeasible solutions.

\begin{obs} \label{obs: all_problems_decidable}
The \PCP is a satisfiability problem in propositional logic. The \CVP falls into the category of difference logic (a fragment of linear arithmetic). Thus, both problems are decidable.
\end{obs}

\begin{obs} \label{obs:problem_is_bounded}
    The \PCP is bounded. In fact, the \PCP involves only a finite number of Boolean variables, so its domain is finite.
\end{obs}

\begin{lemma} \label{obs:finite_number_of_paths}
    Given a finite, directed, weighted graph, the number of paths that connect two arbitrary nodes is finite.
\end{lemma}

\begin{proof}
    By definition, a path is an ordered set of nodes such that no node appears more than once. If the number of nodes in the graph is finite, there cannot be an infinite number of paths. 
\end{proof}

\begin{lemma} \label{lemma:enumerate_all_paths}
    For a given set of routes $\RoutesSet$ and start times in $\Gamma$, repeated calls to the $\naivePathsChanger$ function will enumerate all feasible solutions to the \PCP, either belonging to \feasibleSet or \unfeasibleSet, before returning infeasible. 
\end{lemma}

\begin{proof}
    Let $\varphi_0$ be the conjunction of constraints \eqref{eq:start_and_end_are_true}-\eqref{eq:exactly_two_edges}, a relaxation of the \emph{Paths Changing Problem}, and let $\CurrentPaths_0$ be a solution to $\varphi_0$. Then, if another solution $\CurrentPaths_1$ for $\varphi_0$ exists, it can be found by solving $\varphi_0 \wedge \neg{\CurrentPaths_0} = \varphi_1$. In general, the $n$-th solution can be found by solving $\varphi_0 \wedge \neg{\CurrentPaths_0}\wedge\ldots\wedge\neg{\CurrentPaths_{n-1}} = \varphi_{n}$. Because of \emph{Lemma}~\ref{obs:finite_number_of_paths}, we know that the number of solutions to the \emph{Paths Changing Problem}, $|\solutionSet|$, is finite and we can enumerate them all by solving $\varphi_0,\,\ldots,\,\varphi_{|\solutionSet|-1}$.
\end{proof}

\begin{lemma} \label{lemma:naive_path_search_sound_complete}
    Using the \naivePathsChanger and \emph{CV} is a sound and complete procedure to solve the \conFreePathSearch
\end{lemma}

\begin{proof}
   Because of Observation~\ref{obs:finite_number_of_paths} we know there is a finite number of solutions to the \PCP, and because of Lemma~\ref{lemma:enumerate_all_paths} we know that the \naivePathsChanger function can enumerate them all. If a solution that belongs to \feasibleSet exists the \naivePathsChanger will find it, otherwise it will return all solutions belonging to \unfeasibleSet; the \emph{CV} will then check whether they are conflict-free. Therefore, using the \naivePathsChanger and \emph{CV} in combination will correctly solve the \conFreePathSearch. 
\end{proof}

\begin{lemma} \label{theorem: UCGPathsChanger_does_not_miss_feasible_solutions}
    For a given set of routes $\RoutesSet$, the $\UCGPathsChanger$ is able to find at least all solutions in \feasibleSet.
\end{lemma}

\begin{proof}
    For each set of current paths \CurrentPaths, $\Cbar{\UnsatCore}$ only contains constraints defined by \eqref{eq:nodes_no_swap}, \eqref{eq:edges_direct}, and \eqref{eq:edges_inverse}. 
    The constraints in $\Cbar{\UnsatCore}$ are iteratively retrieved from \emph{minimal} \UC and therefore represent combinations of nodes and edges in the graph where the conflicts happen.
    Since each constraint defined by \eqref{eq:avoid_node} and \eqref{eq:avoid_edge} addresses one constraint from $\Cbar{\UnsatCore}$, \eqref{eq:avoid_node} and \eqref{eq:avoid_edge} only define constraints over nodes or edges that cause conflicts. Hence
    these constraints only remove solutions of the \PCP that belong to \unfeasibleSet.
\end{proof}

\begin{theorem}
     Using the \UCGPathsChanger and \emph{CV} is a sound and complete procedure to solve the \conFreePathSearch.
\end{theorem}

\begin{proof}
    The \naivePathsChanger and the \UCGPathsChanger are identical, except for constraints \eqref{eq:avoid_node}-\eqref{eq:avoid_edge}, and because of Lemma~\ref{theorem: UCGPathsChanger_does_not_miss_feasible_solutions}, we know that the addition of these constraints only removes solutions from \unfeasibleSet. Thus, since the \conFreePathSearch using the \naivePathsChanger is sound and complete (Lemma~\ref{lemma:naive_path_search_sound_complete}), so is the \conFreePathSearch using the \UCGPathsChanger.
\end{proof}

\section{Experiments}\label{sec:experiments}

In order to evaluate the goodness of the proposed method and its performance against the previous version of the \conFreePathSearch algorithm, a set of problem instances is designed and used for testing. Both the \naivePathsChanger and \UCGPathsChanger are embedded in the \theAlgorithm algorithm. However, since the goal is to compare the search for alternative paths, problems are designed in such a way that there is only one feasible set of routes $\RoutesSet$ to serve the customers; also, only the running time for search of conflict-free paths is measured. 
The algorithms called by \theAlgorithm used the SMT solver Z3 4.8.9 to solve the models. All the experiments\footnote{The implementation of the \UCGPathsChanger presented in Section~\ref{sec:improved_paths_search} and the problem instances are available in the \emph{UNSAT\_Core} folder at \url{https://github.com/sabinoroselli/VRP.git}.}
were performed on an \emph{Intel Core i7 6700K, 4.0 GHZ, 32GB RAM}  running \emph{Ubuntu-18.04 LTS}.

Table~\ref{tab:evaluation} shows the results of the evaluation of five problem instances of the \theProblem solved using \theAlgorithm. Each instance was solved twice, once using the 
\naivePathsChanger and once using the \UCGPathsChanger; in each case the number of iterations and the time (in seconds) required to find a feasible solution is reported. The problem instances presented are increasingly hard to solve, in terms of plant size (represented by the number of nodes), number of routes and number of customers in each route. The customers' locations and time windows so that conflicts will arise due to the capacity constraint when the shortest paths are used and a search for alternative paths will be necessary in order to find a conflict-free schedule. 

For instances \emph{1} through \emph{4} it took only one iteration to the \UCGPathsChanger to find a feasible solution, while the \naivePathsChanger required an increasing number of iterations to find a feasible solution, as the instances grew more complicated. The gap in the running time between the \UCGPathsChanger and the \naivePathsChanger follows the same trend; for instance \emph{1} it only takes 2 iterations to the \naivePathsChanger to find a feasible solution, while it takes 24 and 54 iterations to find a solution to instances \emph{2} and \emph{3}. This number drops to 15 iterations for instance \emph{4}. On average, a single iteration of the \naivePathsChanger takes less time than an iteration of the \UCGPathsChanger, but due to the larger number of iterations required, the overall running time for the \naivePathsChanger is always larger.

Instance \emph{5} is the odd one out, as it only takes one iteration of the \naivePathsChanger to find a feasible solution, and, as for the other instances, the running time for the single iteration is shorter. 

\subsection*{Results and Discussion}

The experiments show that for most of the instances the \UCGPathsChanger performed better than \naivePathsChanger in terms of running time and number of iterations. To be more specific, one iteration of the \UCGPathsChanger is slower than one iteration of the \naivePathsChanger, but the number of iterations required by the \naivePathsChanger is always higher, and therefore the overall execution time is longer. As the instances become larger, the gap between the running time for one iteration of each method increases too. However, since the number of iterations required for more complex instances grows as well, the \UCGPathsChanger shows increasing good performance for harder-to-solve instances. On the other hand, Instance \emph{5} shows a different result, since both the \naivePathsChanger and the \UCGPathsChanger take only one iteration. As for the other instances, a single iteration of the \naivePathsChanger is faster, hence the \naivePathsChanger beats the \UCGPathsChanger on Instance \emph{5}. We can conclude that for some instances, the \naivePathsChanger may be able to quickly find feasible solutions and outperform the \UCGPathsChanger. However this is behaviour is highly dependent on the instance and as instances grow larger the chances could grow smaller, as the number of possible paths available increases. Moreover, a detailed analysis of the solutions to the \PCP for each instance\footnote{Details of the problem instances are discussed in the file \emph{Instances\_Results.pdf} in the \emph{UNSAT\_Core} folder of the Github repository.} confirms that, for the \naivePathsChanger, there is no convergence to a feasible solution as the number of iterations increases, since the number of conflicts does not always decrease at the following iteration. On the other hand, the \UCGPathsChanger shows a consistent behaviour as it always takes only one iteration to find feasible solutions.

\begin{table}[htbp]
  \centering
  \caption{Comparison of the \naivePathsChanger and \UCGPathsChanger over a set of instances of the \theProblem. 
            For each instance the number of iterations and the total running time (in seconds) required to find a feasible solution is reported. }
    \begin{tabular}{cccccccc}
    \toprule
    \multirow{2}[4]{*}{Inst.} & \multirow{2}[4]{*}{$|\nodeSet|$} & \multirow{2}[4]{*}{$|\RoutesSet|$} & \multirow{2}[4]{*}{$|\taskSet|$} & \multicolumn{2}{c}{Iterations} & \multicolumn{2}{c}{Time} \\
\cmidrule{5-8}       &    &    &    & \naivePathsChanger & \UCGPathsChanger & \naivePathsChanger & \UCGPathsChanger \\
    \midrule
    1  & 3  & 2  & 4  & 2  & 1  & 0.25 & {\bf 0.16} \\
    2  & 8  & 3  & 6  & 24 & 1  & 8.81 & {\bf 0.40} \\
    3  & 5  & 4  & 8  & 54 & 1  & 35.92 & {\bf 1.08} \\
    4  & 64 & 4  & 28 & 15 & 1  & 643.40 & {\bf 184.60} \\
    5  & 64 & 4 & 28 & 1 & 1 & {\bf 21.20} & 128.40 \\
    \bottomrule
    \end{tabular}%
  \label{tab:evaluation}%
\end{table}%

\section{Conclusions} \label{sec:conclusions}

This paper presents an algorithm to search for conflict-free paths for a set of routes to serve customers in a conflict-free electric vehicle routing problem (\theProblem). The algorithm exploits the SMT solvers' ability to return a \MUC when a formula is infeasible, to guide the search for paths. Soundness and completeness of the algorithm are proved, and preliminary experimental data based on a set of generated \theProblem problem instances are provided. The experiments show that the new \MUC based algorithm consistently finds feasible paths taking only one iteration and significantly shorter time than the previous naive method. Future work includes to run extensive computational analyses to strengthen the claims made in this paper, and further development of the \MUC guided paths search by improving the information extraction from the \MUC.

\balance
\printbibliography{}

\end{document}

%% file: ComSat_Flowchart.tex
\tikzstyle{startstop} = [fill=red!30, ellipse, minimum width=2cm, minimum height=0.5cm,text centered, draw=black]

\tikzstyle{io} = [trapezium, trapezium left angle=70, trapezium right angle=110, minimum width=2cm, minimum height=0.5cm, text centered, draw=black]

\tikzstyle{process} = [fill=cyan!30, rectangle, minimum width=3cm, minimum height=0.5cm, text centered, text width=2cm, draw=black]

\tikzstyle{process1} = [fill=cyan!30, rounded rectangle, minimum width=3cm, minimum height=0.5cm, text centered, text width=2cm, draw=black]

\tikzstyle{decision} = [fill=green!30, diamond, aspect=2, text centered, draw=black]

\tikzstyle{arrow} = [thick,->,>=stealth]
\tikzstyle{edge} = [thick]

\begin{tikzpicture}[node distance=1cm]

\node (start) [startstop] {Start};
\node (router) [process1, below of=start] {Router};
\node (sat1) [decision, below of=router, yshift=-0.5cm] {feasible?};
\node (infeas) [startstop, left of=sat1, xshift=-2cm] {Infeasible};
\node (assign) [process, below of=sat1, yshift=-0.5cm] {Assign};
\node (sat2) [decision, below of=assign, yshift=-0.5cm] {feasible?};
\node (cv) [process, below of=sat2, yshift=-0.5cm] {Capacity Verifier};
\node (sat3) [decision, below of=cv, yshift=-0.5cm] {feasible?};
\node (sch) [startstop, right of=sat3, xshift=2cm] {Schedule};
\node (pc) [process1, below of=sat3, yshift=-0.5cm] {PathsChanger};
\node (sat4) [decision, below of=pc, yshift=-0.5cm] {feasible?};

\coordinate[right of=sat2, xshift=1cm] (coord2);
\coordinate[above of=cv, yshift=-0.2cm] (coord1);
\coordinate[left of=sat4, xshift=-1cm] (coord3);
\coordinate[right of=sat4, xshift=3.5cm] (coord4); 

\node[draw,inner xsep=12pt,dashed,fit={(coord1) (coord3) (coord4) (sat4)}]{};

\node[below of=coord4, yshift=0.6cm, xshift=-1.8cm] {Conflict-free Paths Search};

\draw [arrow] (start) -- (router);
\draw [arrow] (router) -- (sat1);
\draw [arrow] (sat1) -- node[anchor=west,above] {No} (infeas);
\draw [arrow] (sat1) -- node[anchor=south,left] {Yes} (assign);
\draw [arrow] (assign) -- (sat2);
\draw [edge] (sat2) -- node[anchor=east, above] {No} (coord2);
\draw [arrow] (sat2) -- node[anchor=south,left] {Yes} (cv);
\draw [arrow] (coord2) |- (router);
\draw [arrow] (cv) -- (sat3);
\draw [arrow] (sat3) -- node[anchor=east, above] {Yes} (sch);
\draw [arrow] (sat3) -- node[anchor=south,left] {No} (pc);
\draw [arrow] (pc) -- (sat4);
\draw [edge] (sat4) -- node[anchor=east,above] {Yes} (coord4);
\draw [edge] (sat4) -- node[anchor=west,above] {No} (coord3);
\draw [arrow] (coord4) |- (cv);
\draw [arrow] (coord3) |- (assign);

\end{tikzpicture}

%% file: CFPS_Flowchart.tex
\tikzstyle{startstop} = [fill=red!30, ellipse, minimum width=2cm, minimum height=0.5cm,text centered, draw=black]

\tikzstyle{io} = [trapezium, trapezium left angle=70, trapezium right angle=110, minimum width=2cm, minimum height=0.5cm, text centered, draw=black, fill=blue!30]

\tikzstyle{process} = [fill=cyan!30, rectangle, minimum width=3cm, minimum height=0.5cm, text centered, draw=black]

\tikzstyle{opt_process} = [fill=cyan!30, rounded rectangle, minimum width=3cm, minimum height=0.5cm, text centered, draw=black]

\tikzstyle{decision} = [fill=green!30, diamond, aspect=2, text centered, draw=black]

\tikzstyle{arrow} = [thick,->,>=stealth]
\tikzstyle{edge} = [thick]

\begin{tikzpicture}[node distance=1cm]

\node (start) [startstop] {Start};
\node (update_CP_1) [process, below of=start] {$\textrm{CP}\leftarrow \textrm{SP;} \ \textrm{PP.add(CP)}$};
\node (capacity_verifier) [process,below of=update_CP_1] {$\Cbar{\UnsatCore},\textrm{CFS = CapacityVerifier(\RoutesSet,}\,\Gamma\!,\textrm{CP)}$};
\node (sat1) [decision, below of=capacity_verifier, yshift=-0.5cm] {$\textrm{CFS} \neq \emptyset$};
\node (update_CP_2) [process,right of=sat1, xshift=3cm] {$\textrm{CP}\leftarrow \textrm{NP;} \ \textrm{PP.add(NP)}$};
\node (solution) [startstop, left of =sat1, xshift=-2cm]{Schedule};
\node (paths_changer) [opt_process,below of=sat1,yshift=-0.5cm] {$\textrm{NP = GuidedPathsChanger(PP,\,\Cbar{\UnsatCore})}$};
\node (sat2) [decision, below of=paths_changer, yshift=-1cm] {$\textrm{NP} \neq \emptyset$};
\node (no_solution) [startstop, left of =sat2, xshift=-2cm]{Infeasible};

\coordinate[right of=sat2,xshift=3cm] (coord1);
\coordinate[above of=update_CP_2,yshift=0.5cm] (coord2);



\draw [arrow] (start) -- (update_CP_1);
\draw [arrow] (update_CP_1) -- (capacity_verifier);
\draw [arrow] (capacity_verifier) -- (sat1);
\draw [arrow] (sat1) -- node[anchor=west,above] {Yes} (solution);
\draw [arrow] (sat1) -- node[anchor=south,left] {No} (paths_changer);
\draw [arrow] (paths_changer) -- (sat2);
\draw [arrow] (sat2) -- node[anchor=west,above] {No} (no_solution);
\draw [edge] (sat2) -- node[anchor=east,above] {Yes} (coord1);
\draw [arrow] (coord1) -- (update_CP_2);
\draw [edge] (update_CP_2) -- (coord2);
\draw [arrow] (coord2) -- (capacity_verifier);

\end{tikzpicture}